\documentclass[aps,prl,showpacs,floatfix,twocolumn,amsmath,amssymb,preprintnumbers]{revtex4-1}
\usepackage{mathrsfs}
\usepackage[figuresright]{rotating}
\usepackage{amsmath}
\usepackage{amssymb}
\usepackage{graphicx}
\usepackage{color}
\usepackage{dcolumn}
\usepackage{bm}
\usepackage{siunitx}

\usepackage{soul}

\makeatletter

\newcommand{\Rmnum}[1]{\expandafter\@slowromancap\romannumeral #1@}
\makeatother

\begin{document}

\title{Type-I superconductivity in noncentrosymmetric NbGe$_{2}$}

\author{Baijiang Lv$^{1}$}
\author{Miaocong Li$^{1}$}
\author{Jia Chen$^{1}$}
\author{Yusen Yang$^{2}$}
\author{Siqi Wu$^{1}$}
\author{Lei Qiao$^{1}$}
\author{Feihong Guan$^{1}$}
\author{Hui Xing$^{2}$}
\author{Qian Tao$^{1}$}
\author{Guang-Han Cao$^{1,3,5}$}
\author{Zhu-An Xu$^{1,3,4,5}$}\email{zhuan@zju.edu.cn}

\address{$^1$ Zhejiang Province Key Laboratory of Quantum Technology and Device, Department of Physics, Zhejiang University, Hangzhou 310027, China;}
\address{$^2$ Key Laboratory of Artificial Structures and Quantum Control, and Shanghai Center for Complex Physics, School of Physics and Astronomy, Shanghai Jiao Tong University, Shanghai 200240, China;}
\address{$^3$ State Key Laboratory of Silicon Materials, Zhejiang University, Hangzhou 310027, China;}
\address{$^4$ Zhejiang California International NanoSystems Institute, Zhejiang University, Hangzhou 310058, China;}
\address{$^5$ Collaborative Innovation Centre of Advanced Microstructures, Nanjing University, Nanjing 210093, China.}

\date{\today}

\begin{abstract}
Single crystals of NbGe$_{2}$ which crystallize in a
noncentrosymmetric hexagonal structure with chirality are
synthesized and their superconductivity is investigated. Type-I
superconductivity is confirmed by dc magnetization, field-induced
second-to first-order phase transition in specific heat, and a
small Ginzburg-Landau parameter $\kappa_{GL}=0.12$. The
isothermal magnetization measurements show that there is a
crossover from type-I to type-II/1 superconductivity with
decreasing temperature and an unusually enhanced surface
superconducting critical field ($H_{c3}$) is discovered. The band
structure calculations indicate the presence of Kramer-Weyl nodes
near the Fermi level. These observations demonstrate that
NbGe$_{2}$ is an interesting and rare example involving the
possible interplay of type-I superconductivity, noncentrosymmetric
structure and topological properties.
\end{abstract}


\maketitle

\section{\Rmnum{1}. Introduction}

Superconductors with noncentrosymmetric structure (NCS) have stimulated
intensive research attention due to theoretically proposed
possible unconventional pairing
states\cite{smidman2017superconductivity}\cite{Bauer2012Non}. In
noncentrosymmetric crystal structures, the antisymmetric
spin-orbit coupling (ASOC) splits the Fermi surface due to the
electric field gradient in the crystal with broken inversion
symmetry. Cooper pairs that originally belong to the same Fermi
surface would be separated into two different Fermi surfaces.
Large enough ASOC could have a significant effect on the
superconducting state, leading to an admixture of spin-singlet and
spin-triplet pairing states\cite{smidman2017superconductivity}.
Such an admixture of pairing states usually leads to unique
superconducting properties. For example, the heavy fermion
superconductor CePt$_{3}$Si with a noncentrosymmetric structure
has an upper critical field beyond the Pauli
limit\cite{yasuda2004superconducting}, and line nodes are found in
the superconducting gap structure of
Li$_{2}$Pt$_{3}$B\cite{takeya2007specific}. Furthermore,
topological superconductivity is proposed in several
noncentrosymmetric compounds such as
PbTaSe$_{2}$\cite{bian2016topological}, BiPd\cite{sun2015dirac},
and YPtBi\cite{kim2018beyond}.

Meanwhile, according to the Ginzburg-Landau (GL) parameters
$\kappa_{GL}$, superconductors can be categorized as type I with
$\kappa_{GL}<1/\sqrt2$, and type II with $\kappa_{GL}>1/\sqrt2$.
However, when $\kappa_{GL}$ is close to $1/\sqrt2$, there will be
a state between type I and type II, which is called type II/1. In
the type-II/1 superconductors, magnetic flux will enter the
sample, but the flux distribution will be affected by the
attraction interaction between the flux
lines\cite{eilenberger1969structure}. Most superconductors belong to
type II. Only a few are reported to be type I, and most of them
are elementary metals with lower $T_{c}$\cite{roberts1976survey}.
Nevertheless, a few binary and ternary compounds are found to be
type I, for example, ScGa$_{3}$\cite{svanidze2012type},
Al$_{6}$Re\cite{peets2019type}, PdTe$_{2}$\cite{leng2017type},
BeAu\cite{singh2019type},
Rh$_{2}$Ga$_{9}$\cite{shibayama2007superconductivity},
and LiPd$_{2}$Ge\cite{gornicka2020soft}. Interestingly, some type-I
superconductors will become type II/1 at low temperatures,
accompanied by an enhanced surface superconducting
state\cite{kimura2016type, wang2005specific}.

Here, we report our study on the superconductivity in NbGe$_{2}$
which crystallizes in a noncentrosymmetric hexagonal structure.
While the superconductivity in NbGe$_{2}$ was first discovered
back in 1978\cite{remeika1978superconductivity}, the interest in
it has been revived since a report on a possible connection with
its topological band structure in 2018\cite{chang2018topological}.
In this paper, we performed systematic measurements of
resistivity, magnetization and specific heat of single crystalline
NbGe$_{2}$. Our study shows that it is a type-I superconductor
with a crossover from type-I to type-II/1 superconductivity at low
temperatures. An extremely high surface superconducting critical
field ($H_{c3}$) is found, which implies the unconventional nature
of superconductivity, possibly related to its noncentrosymmetric
structure and topological properties.

\section{\Rmnum{2}. Experimental details}

NbGe$_{2}$ single crystals were synthesized by a two-step vapor transport
technique using iodine as the transport agent. High-purity niobium ($99.99\%$, Alfa), and germanium ($99.99\%$,
Alfa) powders were taken in a stoichiometric ratio and mixed in a
glove box under argon atmosphere (the percentage of H$_2$O and
O$_2$ $< 0.1$ \si{ppm}). The mixture was pressed into a pellet,
sealed in an evacuated quatrz tube and then pre-synthesized
at~\SI{1073}{\kelvin} for~$3$ days. The resultant pellet
was ground into powders and mixed with~\SI{100}{mg} of iodine in
a sealed quartz tube. Finally, the tube was placed in a two-zone
furnace at ~\SI{1073}{\kelvin} with a temperature gradient
of~\SI{6}{\kelvin/cm} for~$7$ days. Single
crystals with a typical shape of hexagonal pyramid were obtained
after ultrasonic cleaning in ethanol.

Powder x-ray diffraction (XRD) measurements at room temperature
were carried out on a PANalytical X-ray diffactometer (Model
EMPYREAN) with a monochromatic Cu$K\alpha_1$ radiation and a
graphite monochromator. Lattice parameters were derived by
Rietveld refinement using the program GSAS. The electrical
resistivity and the specific heat was measured in Quantum Design
physical property measurement systems (PPMS-9 and PPMS-14). A
four-probe method was used for the resistivity measurements. The
surface of the sample was polished before making electrical
contacts, and then the gold wires were contacted to the sample by
spot welding. The dc magnetization was measured on a Quantum
Design magnetic property measurement system (MPMS3) equipped with
a $^{3}\rm He$ cryostat. The electronic band structure and density
of states (DOS) were calculated using density functional theory employing
plane-wave basis projected augmented wave (PAW) method as
implemented in Vienna ab initio Simulation Package
(VASP)\cite{kresse1993ab, kresse1999ultrasoft}. After convergence
tests and full structure optimization, the lattice constants and
internal atomic positions from calculation can be well compared
with experimental values within 1\% errorbar  with an energy
cut-off of 520 eV and a $15\times 15\times 9$ $\Gamma$-centered
$k$-point mesh, which is sufficient to converge the total energy to 1
meV/atom.

\section{\Rmnum{3}. Results and discussion}
\begin{figure}[htb]
\centering
\includegraphics[width=8.5cm]{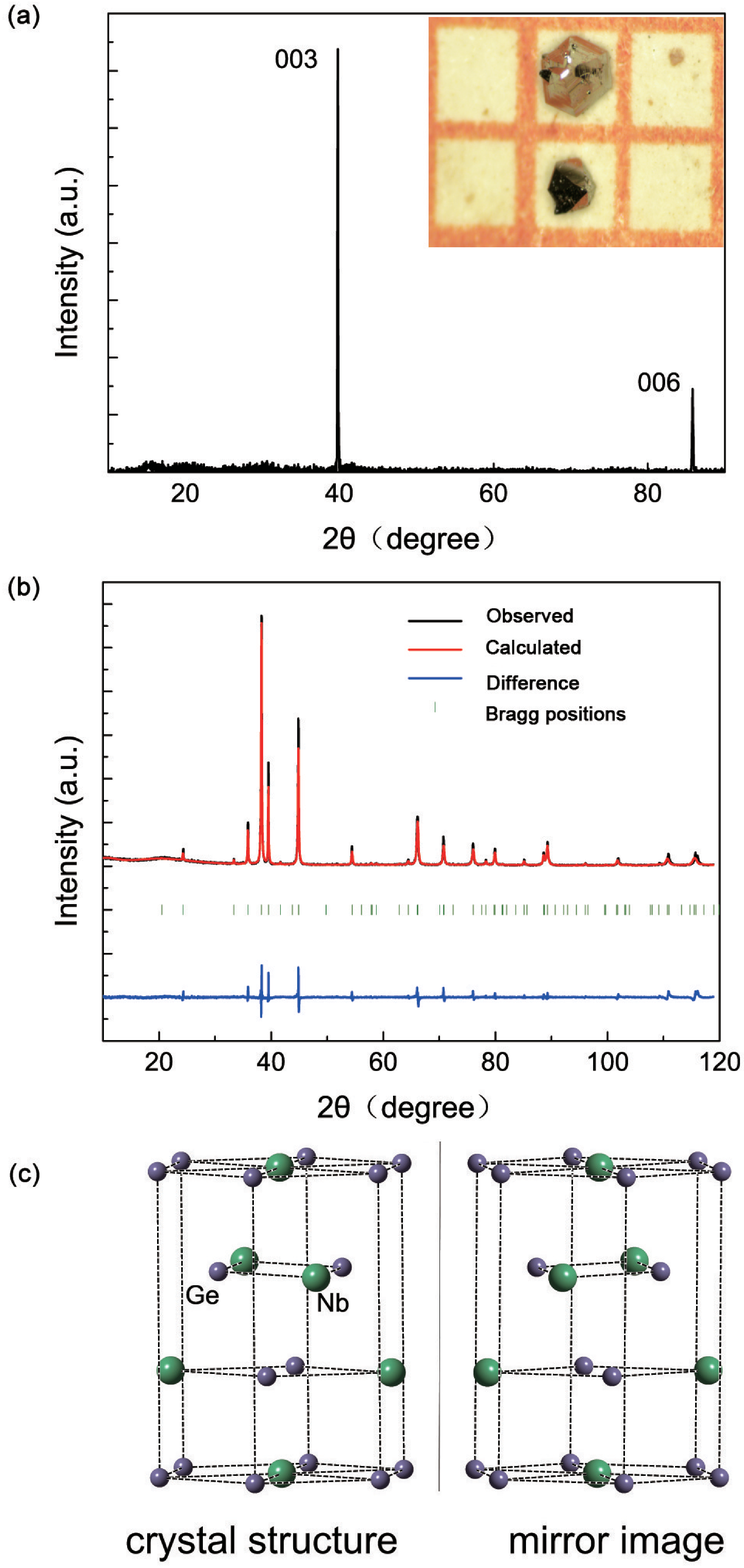}
\caption{(Color online) (a) XRD patterns of a NbGe$_{2}$ single
crystal. The inset shows the typical single crystals used in our
study. (b) Powder XRD patterns of NbGe$_{2}$ single crystals. (c)
The crystal structure of NbGe$_{2}$, which forms in the hexagonal
space group $P6_{2}$22 (No.180) with unit cell parameters $5.002$
\r{A} and $c=6.834$ \r{A}, and its mirror image.} \label{fig1}
\end{figure}

The XRD patterns of a NbGe$_{2}$ single crystal at room temperture
are shown in Fig.1(a). Only sharp (003) and (006) peaks can be
observed, which indicates a uniform $c$-axis orientation
perpendicular to the plane of the single crystal. The inset of
Fig. 1(a) shows a photograph of a typical single-crystalline
sample, which naturally grows into a hexagonal pyramid shape with
a typical size of $1\times1\times1\;\rm mm^3$. The powder XRD
patterns of ground NbGe$_{2}$ single crystals are displayed in
Fig. 1(b). All the patterns can be well indexed with the hexagonal
structure with the space group P$6_{2}$22 (No.180). The refined
lattice parameters are $a=5.002$ \r{A}, $c=6.834$ \r{A} and
$R_{wp}=14.8\%$. The refined atomic positions are summarized in
Table I. The schematic view of NbGe$_{2}$ lattice is shown in Fig.
1(c). The Ge atoms in the top and bottom layers form a honeycomb
lattice, each centered by a Nb atom, and the middle two layers are
alternately arranged by Nb atoms and Ge atoms. The difference
between the middle two layers breaks the inversion symmetry. We
also note that the structure does not overlap with its mirror
image, indicating that its structure is also chiral.

\begin{table}[htpb]
\caption{Refined atomic positions for NbGe$_{2}$ with the space
group P$6_{2}$22 (No.180) and lattice parameters $a=5.002$ \r{A},
$c=6.834$ \r{A}.} \label{tabel:atomic}
\begin{center}
\begin{tabular}{|c|c|c|c|c|c|}
\hline
 Atom & Site & X & Y & Z & Uiso  \tabularnewline
\hline
 Nb  & 3d & 0.500 & 0.000 & 0.500 & 0.00934 \tabularnewline
\hline
 Ge  & 6j & 0.165 & 0.330 & 0.500 & 0.0123 \tabularnewline
\hline
\end{tabular}
\end{center}
\label{default}
\end{table}

The dc susceptibility data measured under a magnetic field of 20
Oe $(H//c)$ in both zero-field-cooled (ZFC) and field-cooled (FC)
modes are shown in the Fig. 2(a). $\chi_{eff}$ has been corrected
by employing the formula: $4\pi\chi_{eff}=4\pi\chi/(1-N\chi)$,
where $N$ is the demagnetization factor and it is about 0.33 in
our case\cite{aharoni1998demagnetizing}. Superconductivity is
observed below the onset point of the diamagnetization signal,
i.e., $T_{c}^{onset} = $ 2.0 $\rm K$. The superconducting volume
fraction estimated from the ZFC data slightly exceeds 100\%,
indicating a good sample quality. Meanwhile, the relatively small
difference between the ZFC and FC curves shows a very small
contribution from magnetic vortices, indicating type-I
superconductivity\cite{svanidze2012type}\cite{peets2019type}.

\begin{figure}[htb]
\centering
\includegraphics[width=8.5cm]{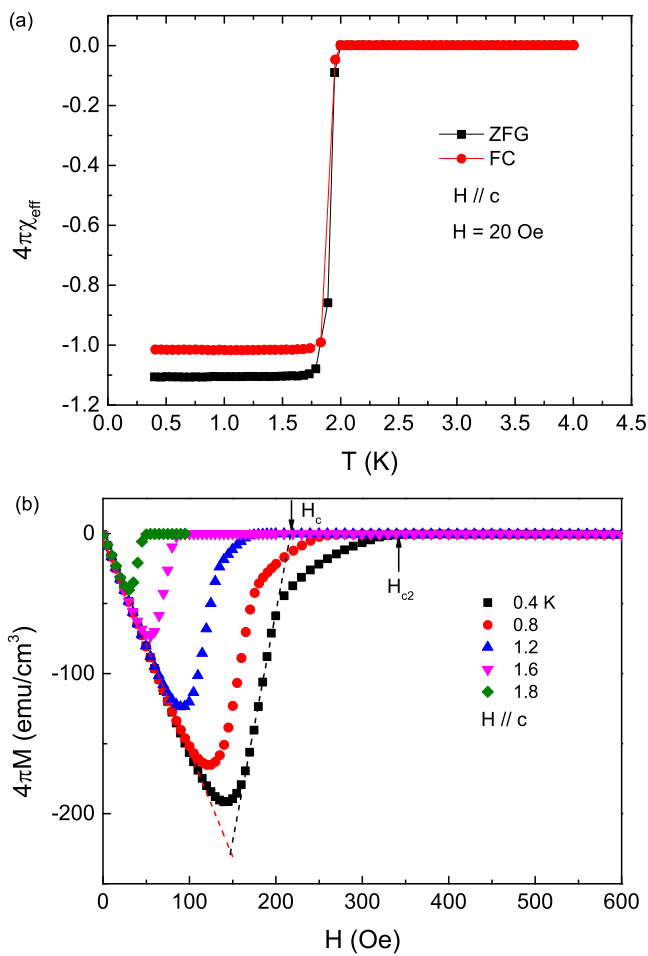}
\caption{(Color online) (a) Temperature dependence of the corrected dc susceptibility
$\chi_{eff}(T)$ in ZFC and FC mode shows the superconductivity at
$T_c^{onset} = 2.0\ \rm K$ in NbGe$_2$. (b) Magnetization $M(H)$ curves at various temperatures.}
\label{fig2}
\end{figure}

The isothermal magnetization $4\pi M (H)$ of NbGe$_{2}$  in the
temperature range $1.8\sim0.4\ \rm K$ is displayed in Fig. 2(b).
We can also get the demagnetization factor $N$ from the initial
slope $-d(4\pi M)/d(H)=1/(1-N)$. The obtained $N$ value is 0.34,
which is consistent with the value evaluated based on the sample
shape. At higher temperatures, e.g., $T$ = 1.6 and 1.8 K, the
$M(H)$ curve exhibits standard type-I behavior: a sharp transition
from the Meissner state to the normal state. For $T<0.4$ K, the
rounding of the $M(H)$ curves is due to the effect of the
demagnetization factor\cite{singh2019type}. We notice that in the
intermediate state the $M(H)$ curve gradually deviates from
linearity, and a tail appears when magnetization approaches zero.
These characteristics are consistent with the reported type-II/1
superconductivity \cite{kimura2016type}\cite{wang2005specific}. In
the type-II/1 superconductivity, the appearance of the small tail
in $M(H)$ is due to the entry of magnetic flux, which leads to a
mixed state. In this case, there is a crossover from type-I to
type-II/1, as temperature decreases. In the type-II/1
superconducting state, there is an attractive interaction between
the flux lines\cite{eilenberger1969structure}. As a consequence, a
Meissner-mixed phase separation state, so called ¡°intermediate
mixed (IM) state¡±, is realized between the Meissner and usual
mixed states\cite{auer1973magnetic}. Such behavior is also
observed in several other type-I
superconductors\cite{kimura2016type}\cite{wang2005specific}. We
can define $H_{c}$ by extending the linear part of the curve to
$M$ = 0 and define $H_{c2}$ as the point where the $M$ reaches 0,
as shown in Fig. 2(b).

\begin{figure}[htb]
\centering
\includegraphics[width=8.5cm]{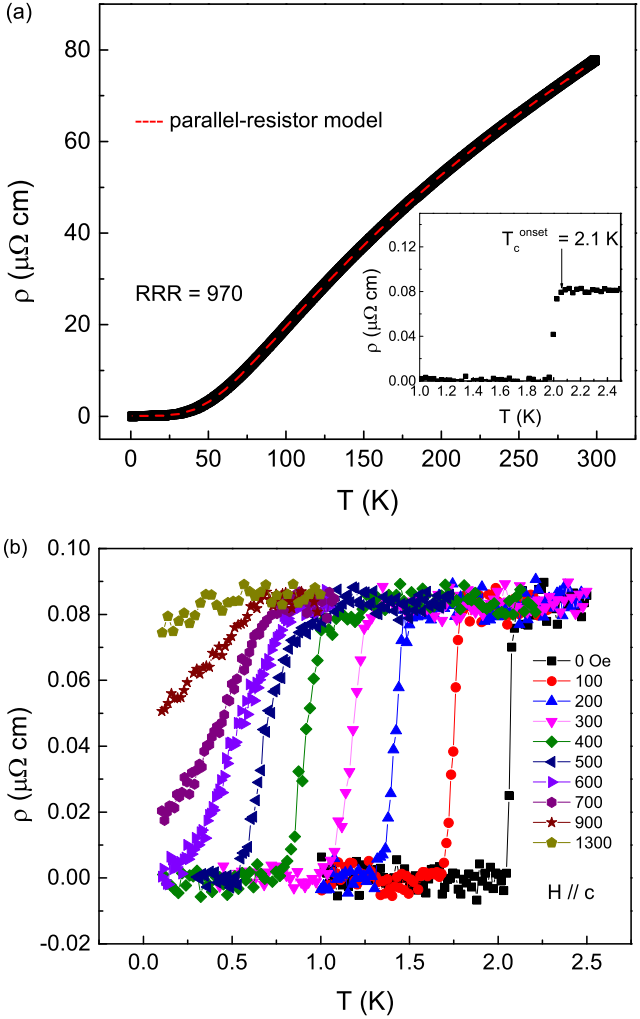}
\caption{(Color online) (a) Temperature dependence of the
electrical resistivity. The red dashed line is a fitting curve of the parallel-resistor model. Inset is an enlarged view of the
superconducting transition in NbGe$_2$. (b) The low-temperature
resistivity (down to 0.1 K) under various magnetic fields.}
\label{fig3}
\end{figure}

Fig. 3(a) shows the temperature dependence of resistivity from 0.5
K to 300 K. The high quality of the sample is clearly demonstrated
by the large ratio of room-temperature resistance to residual
resistance (RRR = $\rho$(300\ K)/$\rho${(2.2\ K)}) , which is up
to 970, much higher than that in earlier report (around
100)\cite{remeika1978superconductivity}. The inset of Fig. 3(a)
shows the superconducting transition around
$T_c^{onset}=\SI{2.1}{\kelvin}$, which is consistent with the
previous reports\cite{remeika1978superconductivity}. In the high
temperature region ($T>100\ K$), the large room temperature
resistivity and a slight negative curvature ($\rm
d^{2}$$\rho/$$\rm d$$T^{2}<0$) was observed. Similar behavior has
been observed in Nb$_3$Sn\cite{fisk1976saturation} and was
interpreted using a parallel resistor model\cite{wiesmann1977simple}. The saturation in
resistivity at high temperatures usually happens when the mean
free path is comparable to the inter-atomic spacing. In general,
the parallel-resistor model can be written as $\frac{1}{\rho
(T)}=\frac{1}{\rho_{1}(T)}+\frac{1}{\rho_{sat}}$, where
$\rho_{sat}$ is the temperature-independent saturation
resistivity, and $\rho_{1}(T)$ is the ideal temperature-dependent
resistivity dominated by electron-phonon scattering:
$\rho_{1}(T)=\rho_{0}+A(\frac{T}{\Theta_{R}})^{5}\int_{0}^{\frac{\Theta_{R}}{T}}\frac{x^{5}dx}{(\exp(x)-1)(1-\exp(-x))}$,
where $\rho_{0}$ denotes residual resistivity, which comes from
impurities and disorder, and the second term is the generalized
Bloch-Gr$\ddot u$neisen expression. A fitting employing this model
is shown in Fig. 3(a), which yields residual resistivity
$\rho_{0}=0.085(0.004)\ \mu\Omega\  \rm cm$, $\rho_{sat}=325.1(3)\
\mu\Omega\  \rm cm$, Debye temperature $\Theta_{R}=335(3)\ \rm K$,
$A=490.4(4)\ \mu\Omega\  \rm cm$. The Debye temperature derived
from the resistivity is comparable with that derived from the
specific heat, $\Theta_{D}=298\ \rm K$ (see below). The saturation
resistivity value is close to that reported in other
compounds\cite{fisk1976saturation}, and we can estimate the mean
free path
$l_{sat}=1.27\times10^{4}\times[\rho_{sat}\times(n^{2/3}\times
S/S_{F})]^{-1}=0.2\ \rm nm$\cite{orlando1979critical}, which is on
the same order of magnitude with inter-atomic spacing.

The low-temperature resistivity measured under various magnetic
fields is shown in Fig. 3(b). The application of a magnetic field
suppresses $T_{c}$ rapidly, but signatures of surviving
superconductivity persists up to a field of 1300 Oe, far exceeding
the critical field obtained in both the magnetization and specific
heat measurements. Since the resistivity is easily dominated by
the surface superconducting states, the above observations may be
related to the contribution from the surface superconductivity,
which will be discussed in more details later.

Fig. 4(a) shows the low temperature specific heat data measured
under zero field. The plot of $C/T$ vs $T^2$ is shown in the inset
of Fig. 4(a). A jump around $T_{c}=\SI{2.0}{\kelvin}$, in
agreement with the observations in the magnetic susceptibility and
resistivity, confirms the bulk superconductivity in NbGe$_2$. The
normal state specific-heat data above $T_{c}$ consists of both
electron and phonon contributions given by $C/T=\gamma_{n}+\beta
T^2$, where $\gamma_{n}$ is the Sommerfeld coefficient and $\beta$
represents the phonon contributions. The red dashed line is the
best fit to the data with $\gamma_{n}=7.35\;\rm mJ/mol\ K^2$,
$\beta=0.22\;\rm mJ/mol\ K^4$ (see the inset in Fig. 4(a)). The
Debye model is then used with the $\beta$ value in the equation
$\Theta_{D}=(12\pi^4NR/5\beta)^{1/3}$ to calculate the Debye
temperature $\Theta_{D}$, where $n=3$  and $R$ is the gas constant
$R=8.31\;\rm J/mol\ K$. The resultant Debye temperature
$\Theta_{D}$ is $\SI{298}{\kelvin}$. With $\Theta_{D}$ and
$T_{c}$, the electron-phonon coupling constant $\lambda_{ep}$ can
then be calculated using the inverted McMillan
equation\cite{mcmillan1968transition}
$\lambda_{ep}=\frac{1.04+\mu^{*}ln(\frac{\Theta_{D}}{1.45T_{c}})}{(1-0.62\mu^{*})ln(\frac{\Theta_{D}}{1.45T_{c}})-1.04}$,
where $\mu^{*}$ is the Coulomb pseudopotential and an empirical
value of $0.13$ is used, and $T_{c}=2\ \rm K$. It yields
$\lambda_{ep}=0.51$, suggesting this material is a weak coupling
superconductor. The DOS at the Fermi level,
$N_{E_{f}}$, is estimated to be  $2.08$ $\rm states/(eV\ f.u.)$
using the relation
$N_{E_{f}}=\frac{3\gamma_{n}}{\pi^2k_{B}^2(1+\lambda_{ep})}$.

\begin{figure}[htb]
\centering
\includegraphics[width=8.5cm]{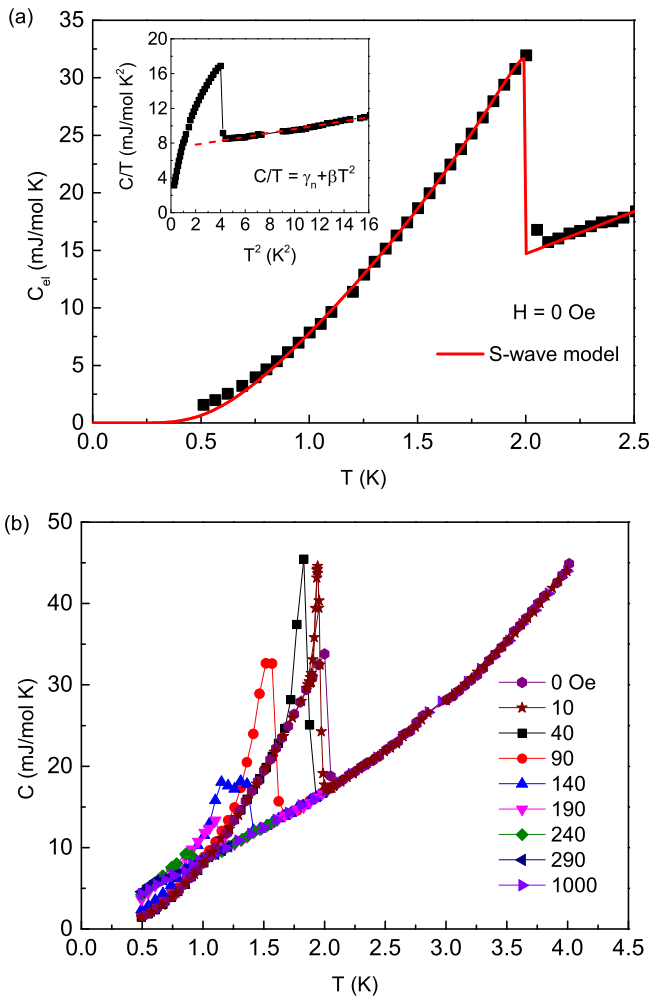}
\caption{(Color online) (a) Temperature dependence of electronic
specific-heat and the fitting with a BCS s-wave model. Inset:
$C/T$ versus $T^2$. (b) Specific heat as a function of temperature
under various magnetic fields.} \label{fig3}
\end{figure}

The electronic contribution ($C_{el}$) to the specific heat
determined by subtracting the phononic contribution from the
measured specific heat data, $C_{el}= C-\beta T^3$, is shown in
the main panel of Fig. 4(a). The value for the specific heat jump,
$\frac{\Delta C_{el}}{\gamma_{n}T_{c}}$, was found to be 1.22,
which is slightly lower than the value for a BCS isotropic gap
superconductor (1.43). This indicates weakly-coupled
superconductivity in NbGe$_2$, which is basically consistent with
the value of $\lambda_{ep}$ obtained above. The temperature
dependence of the specific heat in the superconducting state for a
single BCS gap can be obtained from the normalized entropy $S(T)$
written as: $S(T)=-\frac{6\gamma_{n}}{\pi^{2}
k_{B}}\int_{0}^{\infty}[flnf+(1-f)ln(1-f)]d\epsilon$, where $f =
(e^{\frac{E(\epsilon)}{k_{B}T}}+1)^{-1}$ is the Fermi function,
$E(\epsilon) = \sqrt{\epsilon^{2}+\Delta^{2}(T)}$ is the
excitation energy of quasiparticles, where $\epsilon$ is the
electron energy measured relative to the chemical
potential\cite{tinkham2004introduction,Padamsee1973Quasiparticle},
and $\Delta(T)$ is the temperature dependent gap function, which
in the BCS s-wave model can be approximated as: $\Delta_{s}(T) =
\Delta_{0}\rm tanh[1.82[1.018(T_{c}/T-1)]^{0.51}]$, where
$\Delta_{0}$ is the superconducting gap at zero temperature. The
electronic specific heat is calculated by: $C_{el} =
T\frac{dS}{dT}$. Fitting the specific-heat data using this model
as shown by the solid red line in Fig. 4(a) yields $\alpha =
\Delta_{0}/k_{B}T_{c} = 1.58$, which is less than the BCS value
$\alpha = 1.76$ in the weak-coupling limit. Moreover, we note that
the low-temperature heat capacity slightly deviates from the
universal BCS s-wave model. This is likely due to the
inhomogeneity of the sample which leads to the broadening of the
superconducting transition\cite{peets2019type}, and another
possible reason is that there may be a tiny non-superconducting
impurity phase.

Based on the Sommerfeld coefficient extracted from the specific
heat data, it is possible to estimate the London penetration depth
$\lambda_{L}(0)$, and the coherence length
$\xi(0)$\cite{orlando1979critical}. Since $\rm NbGe_{2}$ has three
formula unit per unit cell, the conduction electron density $n$ is
assumed to be four electrons contributed by Nb. Thus $n=12/V$,
where $V$ is the volume of the unit cell, and we obtain
$n=8\times10^{22}\ \rm cm^{-3}$. If a spherical Fermi surface ($S/S_{F}=1$)
is assumed for this compound, the London penetration depth is
given as
$\lambda_{L}(0)=1.33\times10^{8}\times\gamma^{1/2}\times(n^{2/3}S/S_{F})^{-1}=35.8\
\rm nm$. Meanwhile, the coherence length is determined by using the
BCS relation $\xi(0)=7.95\times10^{-17}\times(n^{2/3}S/S_{F})\times(\gamma
T_{c}^{-1})=295.9\ \rm nm$. The mean free path $l_{tr}$ is estimated
as $l_{tr}=1.27\times10^{4}\times[\rho_{0}\times(n^{2/3}\times S/S_{F})]^{-1}=804.7\
\rm nm$, where $\rho_{0}$ is the low-temperature normal state
resistivity ($0.085\ \mu\Omega\  \rm cm$ at 2.5 K). It clearly
indicates that the electronic mean free path is considerably
larger than the BCS coherence length, and thus the clean limit is
applied to this compound to get the GL
parameter. The GL parameter $\kappa_{GL} =0.957\lambda_{L}(0)/
\xi(0)= 0.12$, which is obviously smaller than
$\frac{1}{\sqrt{2}}$, further confirming that $\rm NbGe_{2}$ is a
type-I superconductor.

The specific heat data under magnetic fields are shown in
Fig. 4(b). A sharp peak is observed under a small
magnetic field of 10 Oe , which signifies a crossover from second to
first-order phase transition. The similar phenomena has been observed
in many type-I superconductors\cite{svanidze2012type, peets2019type}, as a typical feature of type-I superconductivity.
When the magnetic field gradually increases to 50 Oe, the peak magnitude
 also gradually increases. Upon further increasing the magnetic field, the height
of the peak decreases and the peak width broadens.
Far below the transition temperature, the specific heat does not change much with the magnetic
field, implying negligible contributions from the magnetic vortices.

The $H-T$ phase diagram is shown in Fig. 5. $H_{c}$ is determined
by extrapolating the linear part of $M(H)$ curves to $M= 0$ as
shown in Fig. 2(b). $H_{c2}$ is determined by the actual zero
point of $M= 0$ in the $M(H)$ curves, and also by the specific
heat. The $H_{c2}$ values from different methods are consistent
with each other, as shown in Fig. 5. We also define $H_{c3}$ as
the onset point $T_c^{onset}$ in resistivity. $H_{c}$ can be well
fitted by the thermodynamic critical field
formula:$H(T)=H_{c}(0)(1-(\frac{T}{T_{c}})^2)$, which gives $H_{c}(0)
=223\ \rm Oe$. The temperature dependence of $H_{c2}(T)$ was
analyzed by means of  Ginzburg-Landau (GL) model:
$H(T)=H_{c2}(0)\frac{(1-(\frac{T}{T_{c}})^2)}{(1+(\frac{T}{T_{c}})^2)}$,
which gives $H_{c2}(0) =360\ \rm Oe$. The curve of $H_{c3}(T)$ has a
turning temperature $T^{*}$ (shown by the arrow in Fig. 5). Below
$T^{*}$, the curve shows an abnormal divergence at low
temperature, and there is no sign of saturation. Meanwhile, we
note that by defining $T_{c}$ as the midpoint of the resistivity
transition, the main features maintains (shown as blue triangle
and purple cubic in the Fig. 5). $T^{*}$ might be related to the
crossover from type-I to type-II/1 superconductivity, similar to those found in
other type-I compounds\cite{kimura2016type}. If we use linear
fitting of low temperature data, we can get an roughly $H_{c3}(0)$
value equal to 2300 Oe.

\begin{figure}[htb]
\centering
\includegraphics[width=8.5cm]{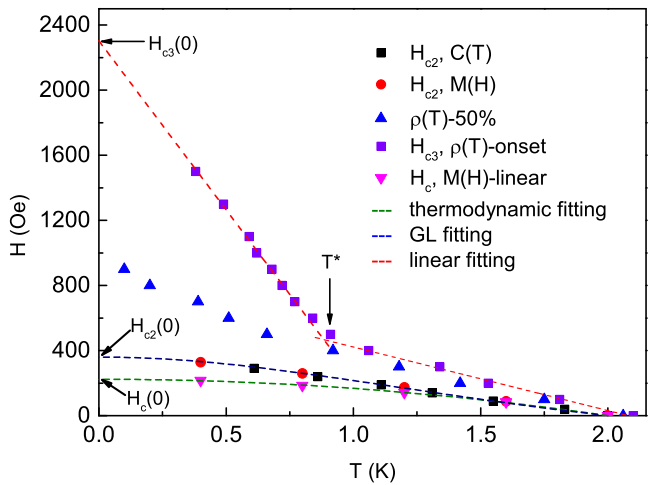}
\caption{(Color online) The $H-T$ phase diagram of the superconductivity of $\rm NbGe_{2}$. $H_{c}$ is estimated by extrapolating the linear part of $M(H)$ curves to $M= 0$, which is fitted by the thermodynamic critical field formula:$H(T)=H_{c}(0)(1-(\frac{T}{T_{c}})^2)$ (green dashed line). $H_{c2}(T)$ is obtained by the actual zero point of $M=0$ in the $M(H)$ curves, or by the specific heat, which is fitted by the GL formula: $H(T)=H_{c2}(0)\frac{(1-(\frac{T}{T_{c}})^2)}{(1+(\frac{T}{T_{c}})^2)}$ (blue dashed line). $H_{c3}$ is defined as the onset point $T_c^{onset}$ in resistivity, and $H_{c3}(0)$ is estimated by linear fitting (red dashed line).}
\label{fig3}
\end{figure}

In the clean limit, superconductivity is known to persist in the
surface up to the surface critical field $H_{c3}\sim 1.7H_{c2}$,
and it is so called standard Saint-James-de Gennes surface
state\cite{Saint1963Onset}\cite{finnemore1966superconducting}. In
our case, the ratio of $H_{c3}/H_{c2}$ equal to 6.4, much larger
than 1.7. We notice that in other type-II/1 superconductors, such
as the centrosymmetric $\rm ZrB_{12}$, the ratio of
$H_{c3}/H_{c2}$ equals to 1.8, which is very close to the
theoretical value\cite{wang2005specific}. While in another
noncentrosymmetric type-II/1 superconductor LaRhSi$_{3}$, $ H_
{c3} / H_ {c2} = 6.7 $, again much larger than the theoretical
value\cite{kimura2016type}. We can see that both NbGe$_{2}$ and
LaRhSi$_{3}$ have an NCS structure, and this high
ratio of $H_{c3}/H_{c2}$ seems related to the NCS structure.
Indeed, in the NCS structure, the ASOC leads to the
surface superconductivity, which has been proposed by the
theoretical
studies\cite{aoyama2014signatures}\cite{iniotakis2008fractional}.
There is also an alternative explanation, i.e., the existence of
topological surface states may also enhance the superconducting
pairing in the surface
states\cite{leng2017type}\cite{liu2015superconductivity}. For
example, in the Dirac semi-metal $\rm PdTe_{2}$, it was found that
the surface superconducting critical field is much higher than the
bulk\cite{leng2017type}. It has been predicted theoretically that
$\rm NbGe_{2}$ can host Kramer-Weyl nodes near the Fermi
level\cite{chang2018topological}. It is interesting to further
explore whether the enhanced surafce critical field is related to
the topological surface states.

\begin{figure}[htb]
\centering
\includegraphics[width=8.5cm]{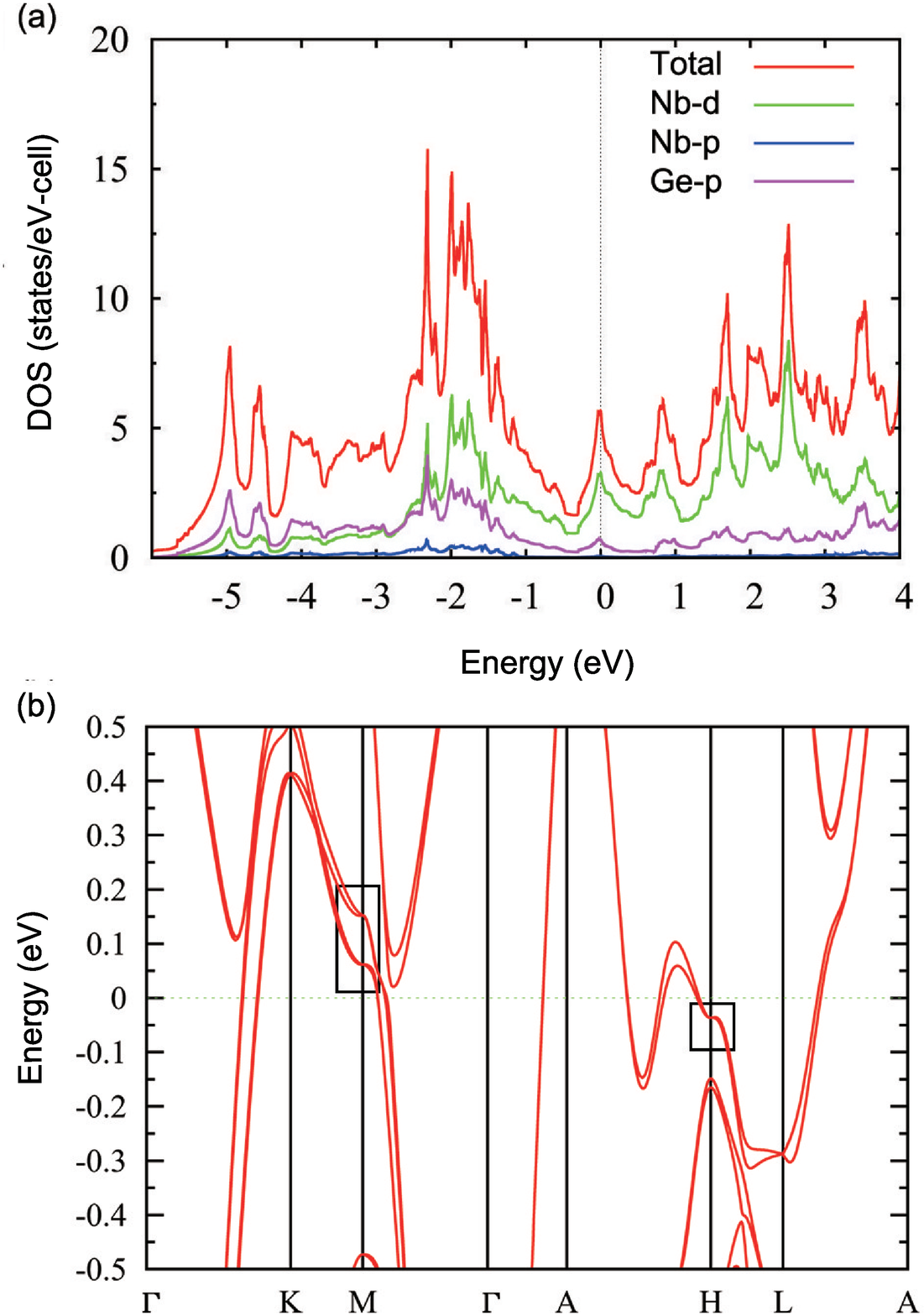}
\caption{(Color online) (a) The calculated DOS
with SOC of NbGe$_{2}$. (b)  Calculated electronic band structure
within $\pm0.5$ eV from the Fermi energy level and Kramer-Weyl
nodes denoted by the black rectangles.} \label{fig3}
\end{figure}

To further understand the properties of superconducting states of
NbGe$_{2}$, density functional calculations of the electronic band
structure with SOC were performed, and the results are displayed
in Fig. 6(a). Near the Fermi level, the DOS is dominated by the Nb
$d$ electrons, and the value equals to 5.64 $\rm states/(eV\
cell)$ $(=1.88\ \rm states/(eV\ f.u.))$, which is comparable to
the experimental value estimated based on the specific heat data.
We can also get the band structure value of the Sommerfeld
coefficient from the density of states $\gamma_{band}=4.4\;\rm
mJ/mol\ K^2$, and the electron-phonon coupling constant
$\lambda_{ep}$ derived from the comparison of $\gamma_{band}$ to
the measured $\gamma_{n}$,
$\lambda_{ep}=\frac{\gamma_{n}}{\gamma_{band}}-1=0.65$, which is
comparable to our previous estimated value based on the McMillan
equation. The finite DOS at a Fermi level also indicates the
metallic ground state, supported by the electrical resistivity
data. The electronic band structure calculated with (SOC) is shown
in Fig. 6(b). The band splitting due to the ASOC is about $\sim$40
meV, which corresponds to a moderately large ASOC effect compared
to other reported
NCS superconductors\cite{smidman2017superconductivity}\cite{nishikayama2007electronic}.
The Kramers-Weyl nodes can be found at the time reversal invariant
points (shown in the black rectangle in the Fig. 6(b)), which is
consistent with the previously reported band
structure\cite{chang2018topological}. Some of them are very close
to the Fermi level, like the nodes near the $M$ point. Meanwhile,
the superconducting state is mainly contributed by the states near
the Fermi level. If these Kramers-Weyl femions participate in the
transport properties, it may have a significant effect on the
superconducting state.

In order to obtain more reliable estimation of the GL parameter,
we should consider the anisotropy of the Fermi velocity in the
calculations\cite{gornicka2020soft}. The anisotropic Fermi
velocity can be obtained from the band structure, and it ranges
from $1.55\times10^{6}$ m/s to $6.95\times10^{6}$ m/s, hence we
can get the BCS coherence length $\xi(0)\cong0.180\frac{\hbar
v_{F}}{k_{B}T_{c}}\cong1.1\times10^{4}-5.0\times10^{4}$ \r{A}.
London penetration depth $\lambda_{L}(0)$ is calculated as
$\lambda_{L}(0)=\sqrt{\frac{3}{\mu_{0}e^{2}v_{F}^{2}N_{E_{f}}}}$,
and we obtained $\lambda_{L}(0)\cong28-126$ \r{A}. We determined
the mean free path using the following equation
\cite{singh2010multigap}
$l=2.372\times10^{-14}\frac{\frac{m^{*}}{m_{e}}^{2}V_{M}^{2}}{N_{E_{f}}^2\rho}$,
where $V_{M}$ is the molar volume, $\rho$ is low-temperature
normal state resistivity, and $N_{E_{f}}$ is the density of states
at the Fermi level. Assuming that$\frac{m^{*}}{m_{e}}=1$, we
obtain $l=6.74\times10^{3}$ \r{A}, which is much smaller than the
BCS coherence length $\xi(0)$, and thus the dirty limit is applied
to this compound to get the GL parameter
$\kappa_{GL} =0.72\lambda_{L}(0)/l(0)\cong3.0\times10^{-3}-1.3\times10^{-2}$. This value of GL
parameter is even smaller than the value calculated based on the
spherical Fermi surface, further supporting the type-I
superconductivity in NbGe$_{2}$.

\section{\Rmnum{4}. Conclusion}
In summary, we have synthesized NbGe$_{2}$ single crystals with
high quality. Based on the resistivity, magnetization and specific
heat measurements, NbGe$_{2}$ is characterized as a type-I BCS
superconductor, and there is a crossover from type-I to type-II/1
superconductivity upon decreasing temperature. A surface
superconducting critical field ($H_{c3}$) much larger than the
bulk one is discovered and we propose that both the
noncentrosymmetric structure and the topological state may be
responsible for such behavior. NbGe$_{2}$ provides a rare example
to explore the possible interplay of type-I superconductivity,
noncentrosymmetric structure and topological surface states.

\section*{Acknowledgments}
We thank Huiqiu Yuan, Yang Liu, and Xin Lu for insightful
discussions. This work was supported by the National Key R\&D
Program of the China (Grant No. 2016YFA0300402, and
2019YFA0308602), the National Science Foundation of China (Grant
Nos. 11774305) and the Fundamental Research Funds for the Central
Universities of China. DFT calculations were performed at the High
Performance Computing Center of College of Science at Hangzhou
Normal University.


\begin{thebibliography}{}
\bibitem{smidman2017superconductivity}M. Smidman, M. Salamon, H. Yuan, and D. Agterberg, {\rm Rep. Prog. Phys.\/} {\bf 80}, 036501 (2017).
\bibitem{Bauer2012Non}E. Bauer and M. Sigrist, {\em Non-centrosymmetric Superconductor: Introduction and Overview\/} (Springer-Verlag, Heidelberg,
2012).
\bibitem{yasuda2004superconducting}T. Yasuda, H. Shishido, T. Ueda, S. Hashimoto, R. Settai, T. Takeuchi, T. D Matsuda, Y. Haga, and Y. {\=O}nuki, {\rm J. Phys. Soc. Jpn.\/} {\bf 73}, 1657 (2004).
\bibitem{takeya2007specific}H. Takeya, M. ElMassalami, S. Kasahara, and K. Hirata, {\rm Phys. Rev. B\/} {\bf 76}, 104506 (2007).
\bibitem{bian2016topological}G. Bian, T.-R. Chang, R. Sankar, S.-Y Xu, H. Zheng, T. Neupert, C.-K. Chiu, S.-M Huang, G. Chang, I. Belopolski, et al., {\rm Nat. Commun.\/} {\bf 7}, 6633 (2016).
\bibitem{sun2015dirac}Z. Sun, M. Enayat, A. Maldonado, C. Lithgow, E. Yelland, D. C. Peets, A. Yaresko, A. P. Schnyder, and P, Wahl, {\rm Nat. Commun.\/} {\bf 6},  (2015).
\bibitem{kim2018beyond}H. Kim, K. Wang, Y. Nakajima, R. Hu, S. Ziemak, P. Syers, L. Wang, H. Hodovanets, J. D. Denlinger, P. M. Brydon, et al., {\rm Sci. Adv.\/} {\bf 4}, eaao4513 (2018).
\bibitem{eilenberger1969structure}G. Eilenberger, and H. B\"uttner, {\rm Z. Physik\/} {\bf 224}, 335 (1969).

\bibitem{roberts1976survey}B. W. Roberts, {\rm J. Phys. Chem. Ref. Data\/} {\bf 5}, 581 (1976).
\bibitem{svanidze2012type}E. Svanidze, and E. Morosan, {\rm Phys. Rev. B\/} {\bf 85}, 174514 (2012).
\bibitem{peets2019type}D. C. Peets, E. Cheng, T. Ying, M. Kriener, X. Shen, S. Li, and D. Feng, {\rm Phys. Rev. B\/} {\bf 99}, 144519 (2019).
\bibitem{leng2017type}H. Leng, C. Paulsen, Y. K. Huang, and A. de Visser, {\rm Phys. Rev. B\/} {\bf 96}, 220506(R) (2017).
\bibitem{singh2019type}D. Singh, A. D. Hillier, and R. P. Singh, {\rm Phys. Rev. B\/} {\bf 99}, 134509 (2019).
\bibitem{shibayama2007superconductivity}T. Shibayama, M. Nohara, H. A. Katori, Y. Okamoto,Z. Hiroi, and H. Takagi, {\rm J. Phys. Soc. Jpn.\/} {\bf 76}, 073708 (2007).

\bibitem{gornicka2020soft}K. G\'ornicka, G. Kuderowicz, E. M. Carnicom, K. Kutorasi\'nski, B. Wiendlocha, R. J. Cava, and T. Klimczuk, {\rm Phys. Rev. B\/} {\bf 102}, 024507 (2020).
\bibitem{kimura2016type}N. Kimura, N. Kabeya, K. Saitoh, K. Satoh, H. Ogi, K. Ohsaki, and H. Aoki, {\rm J. Phys. Soc. Jpn.\/} {\bf 85}, 024715 (2016).




\bibitem{wang2005specific}Y. Wang, R. Lortz, Y. Paderno, V. Filippov, S. Abe, U. Tutsch, and A. Junod, {\rm Phys. Rev. B\/} {\bf 72}, 024548 (2005).
\bibitem{remeika1978superconductivity}J. Remeika, A. Cooper, Z. Fisk, and D. Johnston, {\rm J. Less Common Met\/} {\bf 62}, 211 (1978).
\bibitem{chang2018topological}G. Chang, B. J. Wieder, F. Schindler, D. S. Sanchez, I. Belopolski, S.-M Huang, B. Singh, D. Wu, T.-R Chang, T. Neupert, et al., {\rm Nat. Mater.\/} {\bf 17}, 978 (2018).
\bibitem{kresse1993ab}G. Kresse, and J. Hafner, {\rm Phys. Rev. B\/} {\bf 47}, 558 (1993).
\bibitem{kresse1999ultrasoft}G. Kresse, and D. Joubert, {\rm Phys. Rev. B\/} {\bf 59}, 1758 (1999).
\bibitem{aharoni1998demagnetizing}A. Aharoni, {\rm J. Appl. Phys.\/} {\bf 83}, 3432 (1998).
\bibitem{auer1973magnetic}J. Auer, and H. Ullmaier, {\rm Phys. Rev. B\/} {\bf 7}, 136 (1973).



\bibitem{fisk1976saturation}Z. Fisk, and G. Webb, {\rm Phys. Rev. Lett.\/} {\bf 36}, 1084 (1976).
\bibitem{wiesmann1977simple}H. Wiesmann, M. Gurvitch, H. Lutz, A. Ghosh, B. Schwarz, M. Strongin, P. Allen, and J. Halley, {\rm Phys. Rev. Lett.\/} {\bf 38}, 782 (1977).
\bibitem{orlando1979critical}T. P. Orlando, E. J. McNiff Jr, S. Foner, and M. R. Beasley, {\rm Phys. Rev. B\/} {\bf 19}, 4545 (1979).
\bibitem{mcmillan1968transition}W. McMillan, {\rm Phys. Rev.\/} {\bf 167}, 331 (1968).
\bibitem{tinkham2004introduction}M. Tinkham, {\em Introduction to Superconductivity}, 2nd ed. (Dover Publications, Mineola, NY, 1996).
\bibitem{Padamsee1973Quasiparticle}H. Padamsee, J. E. Neighbor, and C. A. Shiffman, {\rm J. Low temp. Phys.\/} {\bf 12}, 387 (1973).
\bibitem{Saint1963Onset}D. Saint-James, and P. G. Gennes, {\rm Phys. Lett.\/} {\bf 7}, 306 (1963).
\bibitem{finnemore1966superconducting}D. Finnemore, T. Stromberg, and C. Swenson, {\rm Phys. Rev.\/} {\bf 149}, 231 (1966).
\bibitem{aoyama2014signatures}K. Aoyama, L. Savary, and M. Sigrist, {\rm Phys. Rev. B\/} {\bf 89}, 174518 (2014).
\bibitem{iniotakis2008fractional}C. Iniotakis, S. Fujimoto, and M. Sigrist, {\rm J. Phys. Soc. Jpn.\/} {\bf 77}, 083701 (2008).
\bibitem{liu2015superconductivity}Z. Liu, X. Yao, J. Shao, M. Zuo, L. Pi, S. Tan, C. Zhang, and Y. Zhang, {\rm J. Am. Chem. Soc.\/} {\bf 137}, 10512 (2015).
\bibitem{nishikayama2007electronic}Y. Nishikayama, T. Shishidou, and T. Oguchi, {\rm J. Phys. Soc. Jpn.\/} {\bf 76}, 064714 (2007).
\bibitem{singh2010multigap}Y. Singh, C. Martin, S. L. Bud¡¯ko, A. Ellern, R. Prozorov, and D. C. Johnston, {\rm Phys. Rev. B\/} {\bf 82}, 144532 (2010).




\end{thebibliography}
\end{document}